\documentclass[preprint,12pt]{elsarticle}

\usepackage{amssymb}
\usepackage{amsthm}
\usepackage{lineno}

\journal{Nucl. Instrum. Methods Phys. Res. A}

\begin{document}

\begin{frontmatter}



\title{Study on the anti-correlated painting injection scheme for the Rapid Cycling Synchrotron of the China Spallation Neutron Source}


\author[address1,address2,address3]{Ming-Yang Huang}
\author[address1,address3]{Shouyan Xu}
\author[address1,address3]{Yuwen An}
\author[address1,address3]{Jianliang Chen}
\author[address1,address3]{Liangsheng Huang}
\author[address1,address3]{Mingtao Li}
\author[address1,address3]{Yong Li}
\author[address1,address3]{Zhiping Li}
\author[address1,address3]{Xiaohan Lu}
\author[address1,address3]{Jun Peng}
\author[address1,address3]{Yue Yuan}
\author[address1,address2,address3]{Sheng Wang\fnref{footnote1}}
\fntext[footnote1]{Corresponding author email: wangs@ihep.ac.cn}

\address[address1]{Institute of High Energy Physics, Chinese Academy of Sciences, Beijing 100049, China}
\address[address2]{University of Chinese Academy of Sciences, Beijing 100049, China}
\address[address3]{Spallation Neutron Source Science Center, Dongguan 523808, China}

\begin{abstract}

In the rapid cycling synchrotron of the China Spallation Neutron Source, the anti-correlated painting was adopted for the design scheme
of the injection system. In the beam commissioning, with the optimization of the anti-correlated painting,
the injection beam loss has been well controlled and the injection efficiency has exceeded 99\%. Combined with other
aspects of adjustments, the beam power on the target has reached 50 kW smoothly.
In this paper, we have studied the injection optimization in the beam commissioning.
Compared to the simulation results of the design scheme,
the transverse beam distribution, transverse coupling effect and beam loss of the anti-correlated painting in the beam commissioning
are somewhat different.
Through the machine studies, we have carefully analyzed these differences and studied their reasons.

\end{abstract}



\begin{keyword}
CSNS \sep Injection \sep Painting \sep Beam commissioning

\PACS  29.25.Dz \sep 29.27.-a \sep 29.27.Ac
\end{keyword}

\end{frontmatter}


\section{Introduction}
\label{Int}

The China Spallation Neutron Source (CSNS) is a high power proton accelerator-based
facility with the design beam power of 100 kW and repetition rate of 25 Hz \cite{Wang1}\cite{Wei1}.
The schematic layout of the CSNS is shown in Fig. 1.
Its accelerator consists of a 1.6 GeV rapid cycling
synchrotron (RCS) and an 80 MeV H$^-$ Linac \cite{Wei2}. The RCS accumulates
and accelerates an 80 MeV injection
beam to the design energy of 1.6 GeV and then
extracts the high energy beam to the target.
Its lattice has a four-fold structure with four long straight sections for the injection,
extraction, RF cavity and beam collimation respectively.

\begin{figure}
\centering\includegraphics[width=0.9\textwidth]{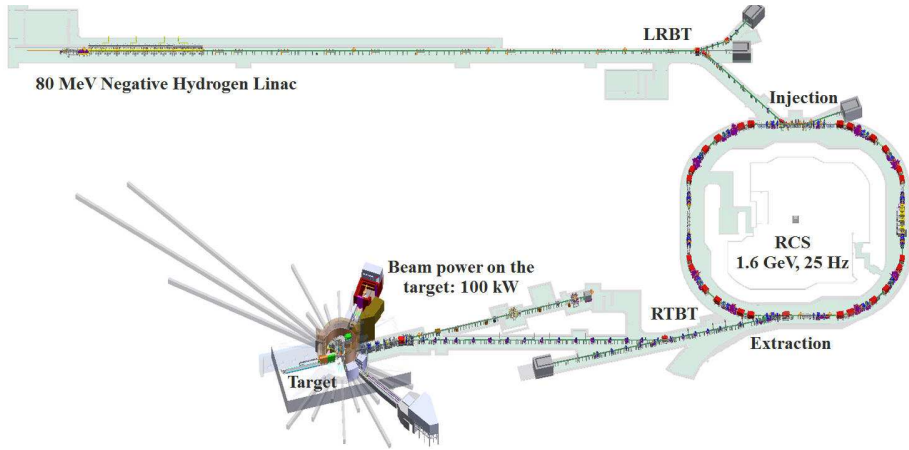}\\
\caption{The schematic layout of the CSNS.}
\end{figure}

The injection system is the core part of the CSNS accelerator and the injection efficiency
is an important factor that determines whether the accelerator can operate safely \cite{Huang1}.
The injection process determines the initial state of the circulating beam and has
an important influence on the processes of beam accumulation and acceleration. For the injection system,
the H$^-$ stripping \cite{Kawakubo1}\cite{Mohagheghi1}\cite{Drozhdin1} is adopted to
inject the Linac beam to the RCS with high precision and high efficiency.
In order to control the strong space charge effects \cite{Holmes1}\cite{Cousineau1}\cite{Xu1}, which is the main cause
of the beam loss, the
phase space painting is used \cite{Wei3}\cite{Beebe-Wang1}\cite{Beebe-Wang2}
for injecting a small emittance beam from the Linac into the large acceptance of the RCS.
The anti-correlated painting scheme was adopted as the design scheme for the
injection system \cite{Huang1}\cite{Tang1}.

The paper is organized as follows. In Sec. 2, the design scheme of the anti-correlated
painting is introduced and discussed.
In Sec. 3, the injection beam commissioning is studied.
The results of the beam commissioning are given and compared to the simulation results of the design scheme.
The summary and discussion are given in the last section.

\section{Design scheme of the anti-correlated painting}
\label{SecII}

In the design stage, by using the codes ORBIT \cite{Gabambos1} and SIMPSONS \cite{Machida1},
the correlated painting and anti-correlated
painting of the injection process and acceleration process were simulated \cite{Qiu1}\cite{weit1}\cite{Xu2}.
According to the simulation results, compared to the correlated painting,
the transverse coupling effect and beam loss of the anti-correlated painting
with the nominal working point (4.86, 4.78) were smaller.
Therefore, the anti-correlated painting scheme was adopted.

For the CSNS/RCS, the injection system consists of four
bump magnets BC1-BC4 which generates a horizontal chicane bump of 60 mm,
four horizontal painting magnets BH1-BH4, four vertical painting magnets BV1-BV4,
two septum magnets (ISEP1, ISEP2), a primary stripping foil (Str-1),
and a secondary stripping foil (Str-2), as shown in Fig. 2.

\begin{figure}
\centering\includegraphics[width=0.7\textwidth]{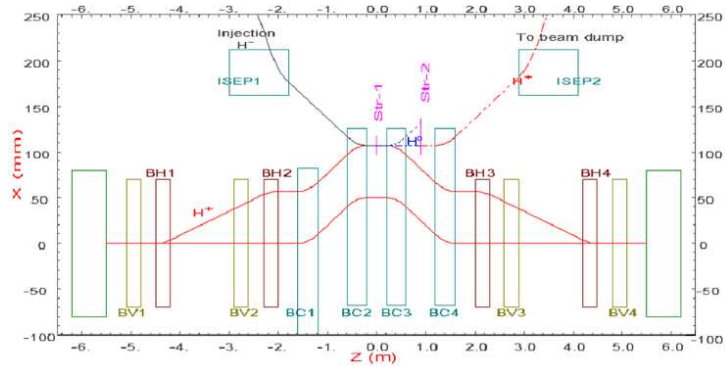}\\
\caption{The schematic diagram of the injection system.}
\end{figure}

In the injection process,
the circulating beam moves from the positive maximum to the center in the horizontal plane while
it moves from the negative maximum to the center in the vertical plane. The injection point is located in
the lower left corner of the main stripping foil to reduce the traversal times. Fig. 3 shows the schematic of
the RCS acceptance ellipse and injection beam in the injection process.

\begin{figure}
\centering\includegraphics[width=0.95\textwidth]{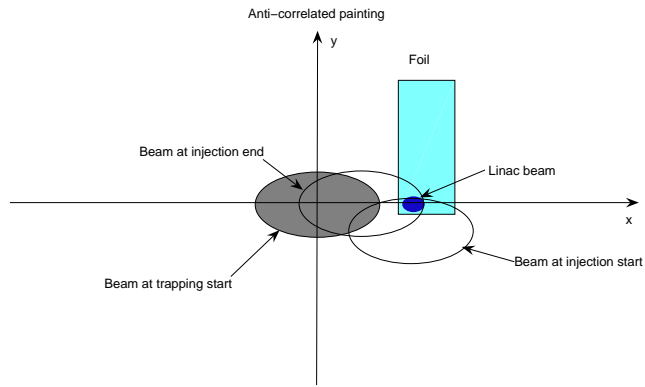}\\
\caption{Schematic of the RCS acceptance ellipse and injection beam in the anti-correlated painting process.}
\end{figure}

\begin{figure}
\centering\includegraphics[width=0.6\textwidth]{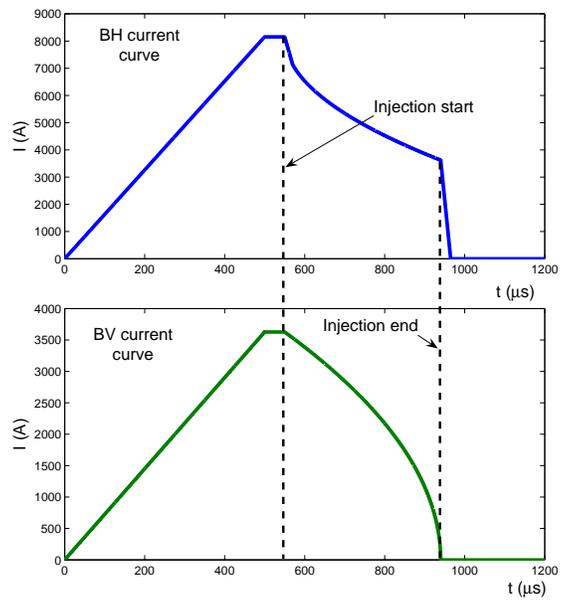}\\
\caption{The BH and BV pulse current curves for the anti-correlated painting scheme.}
\end{figure}

For the anti-correlated painting, a square-root-type function for
the phase-space offset of the injection beam relative to the RCS closed orbit is chosen:
\begin{eqnarray}
X=X_{max}-(X_{max}-X_{min})\sqrt{\frac{t}{T_{inj}}}, \quad X'= 0, \label{CX}
\end{eqnarray}
in the horizontal plane and
\begin{eqnarray}
Y=-Y_{max}\times\sqrt{1-\frac{t}{T_{inj}}}, \quad Y'= 0, \label{CY}
\end{eqnarray}
in the vertical plane, where $X_{max} - X_{min}$ is the horizontal painting range,
$Y_{max}$ is the vertical painting range,
$T_{inj}$ is the injection time which is 0.39 ms. Since the injection
point of the Linac beam sets at ($X_{max}$, 0), the
maximum horizontal painting emittance depends on $|X_{max}-X_{min}|$ and the maximum vertical
painting emittance depends on $|Y_{max}|$. Fig. 4 shows the BH and BV pulse current
curves for the anti-correlated painting scheme in which the falling edges
of the two current curves are both used for painting.

In order to describe the basic idea of the painting injection in terms of the
painting area size in comparison with the machine aperture along the injection bump orbit,
the relationship between the emittance of the circulating beam relative to the vacuum chamber center
and the machine aperture of the vacuum chamber needs to be studied in detail.
Since the dispersion functions $D_x\sim 0$ and $D_y\sim 0$ near the injection point, the beam size of the circulating beam can be calculated by
\begin{eqnarray}
\sigma_{x}^2=\beta_x\cdot\varepsilon_x, \quad \sigma_{y}^2=\beta_y\cdot\varepsilon_y,  \label{sigmaxy}
\end{eqnarray}
where $\varepsilon_x$ and $\varepsilon_y$ are the horizontal and vertical emittances of the circulating beam,
$\beta_x$ and $\beta_x$ are the Courant-Snyder parameters near the injection point. Supposing the deviation position
of the vacuum chamber center relative to circulating beam as ($x_0$, $y_0$), the horizontal
and vertical emittances of the circulating beam relative to
the vacuum chamber center can be calculated by
\begin{eqnarray}
\varepsilon_{rx}=\frac{[|X-x_0|+|\sigma_x|]^2}{\beta_x}, \quad \varepsilon_{ry}=\frac{[|Y-y_0|+|\sigma_y|]^2}{\beta_y}, \label{erxy}
\end{eqnarray}
where $X$ and $Y$ are the horizontal and vertical painting positions during the injection process.
Therefore, supposing the machine aperture radius of the circular vacuum chamber as $R$, the injection beam loss requires
that the circulating beam should meet the following condition in the whole painting process:

\begin{eqnarray}
R > \sqrt{\varepsilon_{rx}\cdot \beta_x + \varepsilon_{ry}\cdot \beta_y}. \label{kxy}
\end{eqnarray}

\section{Beam commissioning}
\label{SecIII}

In this section, the results and problems of the beam commissioning for the CSNS injection system
will be studied. First of all, the beam parameters matching between the Linac beam and
circulating beam should be studied. In order to complete the beam accumulation process as soon as possible,
the fixed-point injection method was used in the early stage of the beam commissioning. The fixed-point injection method means
that the relative positions of the injected beam center and circulating beam center remain constant
during the injection process. With the increase of injection beam power,
the transverse phase space painting was used to reduce the beam loss.
The painting ranges and painting curves were optimized and compared to the simulation results of the design scheme.
In order to solve the problem of the middle concave of the extraction beam distribution, three aspects of
adjustments were carried out. After the above optimizations,
the injection beam loss has been well controlled and the injection efficiency has been over 99\%.
Combined with other aspects of the beam commissioning, the beam power on the target has been
over 50 kW and the stable operation can be achieved.

\subsection{Injection beam parameters matching}
\label{SecIII1}

A mismatch between the injection beam and circulating beam can result in large beam loss
and an undesirable transverse emittance growth. There are three aspects which should be considered
in the injection beam parameters matching: twiss parameters, dispersion function, and beam orbit.
The twiss parameters matching is achieved by
\begin{eqnarray}
 \frac{\alpha_l}{\beta_l}=\frac{\alpha_r}{\beta_r},
\label{match}
\end{eqnarray}
where $\alpha_l$ and $\beta_l$ are the twiss parameters for the Linac,
$\alpha_r$ and $\beta_r$ are that for the RCS. For CSNS, the nominal $\alpha_r$ equals to 0.
The simulation results in Ref. \cite{Huang1} show that the effects of the twiss parameters mismatch are
very small when $\alpha_l<1$. In the beam commissioning, it can be
proved that the injected beam state was almost unchanged when $\alpha_l<1$ which can be easy to achieve
by adjusting the magnets on the beam line from the Linac to the RCS (LRBT).
Therefore, the requirement of the injection
twiss parameters matching is easy to satisfy.
For CSNS, the nominal dispersion functions in the injection region for both injection beam and circulating beam
are 0. In the beam commissioning, it can be found that the actual dispersion functions in the injection region
for both injection beam and circulating beam were very small which can be neglected.
In the following sections, we will focus on the methods of the
orbit matching between the injection beam and circulating beam.

\begin{figure}
\centering\includegraphics[width=1.0\textwidth]{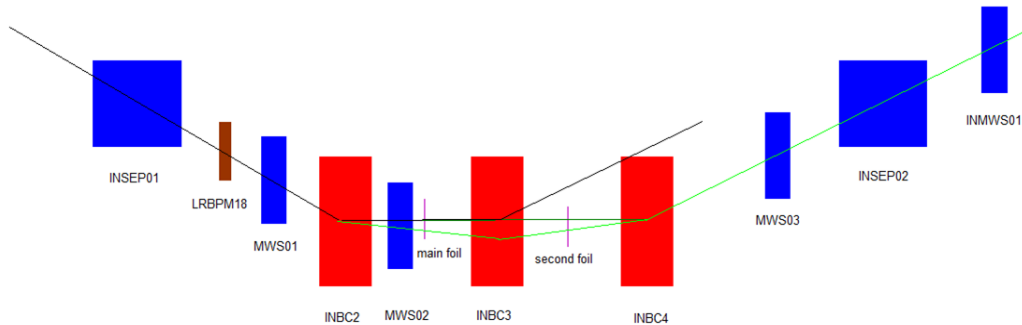}\\
\caption{The schematic layout of the I-dump beam line.}
\end{figure}

In order to measure and adjust the injection beam orbit precisely,
the I-dump beam line mode was designed \cite{Huang2}, as shown in Fig. 5.
There are four multi-wire scanners (MWs) on the I-dump beam line.
In the beam commissioning, by using two of the four MWs, the phase space coordinates of the Linac beam at
the injection point can be measured. By adjusting the injection magnets,
the phase space coordinates of the injection beam can be calibrated to the reasonable
ranges which can basically meet the requirements of the injection beam orbit matching.

In Ref. \cite{Lu1}, a new and more accurate method based on multi-turn injection and Fourier fitting was developed to
match the injection beam orbit. The machine study shows that the mismatch of injection
parameters can be well depressed by using this new method.

\subsection{A comparative study of the fixed-point injection and painting injection}
\label{SecIII2}

\begin{figure}
\centering\includegraphics[width=0.7\textwidth]{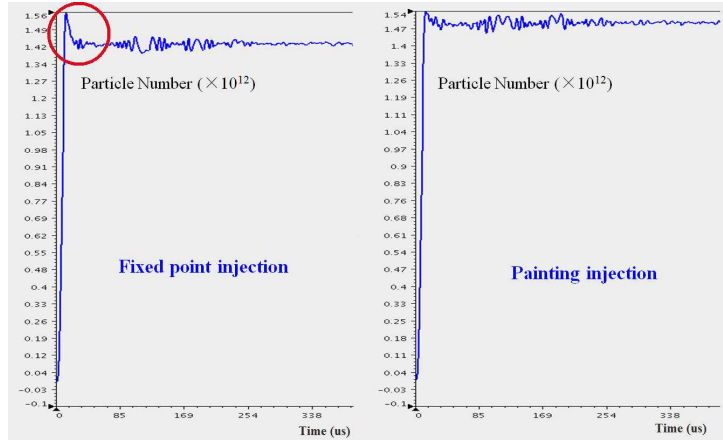}\\
\caption{RCS DCCT displays when the fixed-point injection and painting injection were used.}
\end{figure}

Since the injection beam power was relatively small in
the early stage of the beam commissioning, the fixed-point injection method was selected.
However, when the beam power on the target reached 10 kW,
there was a sudden beam loss in the injection
process no matter the matching between the injection beam and circulating beam, as shown in Fig. 6.
By using the transverse phase space painting instead of the fixed-point injection,
the sudden beam loss in the
injection process was gone. Fig. 6 shows the RCS direct-current current transformer (DCCT)
display when the fixed-point injection and painting injection were used.
It can be found that, with the transverse phase space painting, the sudden beam loss was gone
and the injection efficiency had been improved. The main reason of this
phenomenon is that, on the one hand, when the beam density increase in the case of the fixed-point
injection, the total emittance of the circulating beam will increase caused by the strong space
charge effects; on the other hand, in the fixed-point injection, to match the injection beam,
the center of the circulating beam is bumped to the wall of the vacuum chamber, and the acceptance
of injection area is accordingly decreased. In the case of the painting injection, the space
charge effects will be depressed by painting, and the center of the circulating beam
will move to the center of the vacuum chamber during the injection process.

\subsection{Study on the painting ranges and transverse coupling effect}
\label{SecIII3}

\begin{figure}
\centering\includegraphics[width=1.0\textwidth]{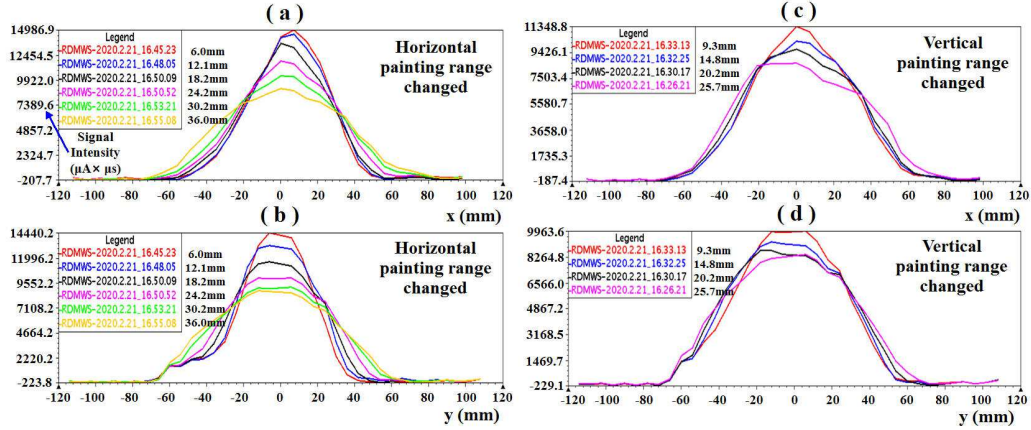}\\
\caption{Transverse beam distributions changing with the horizontal and vertical painting ranges when the anti-correlated painting
and working point (4.81, 4.87) were used. Subgraphs (a) and (b): the horizontal painting range was changed when the vertical
painting range was fixed to 12 mm; subgraphs (c) and (d): the vertical painting range was changed when the horizontal
painting range was fixed to 27 mm.}
\end{figure}

According to the simulation results of the design scheme,
the optimal horizontal and vertical painting ranges are 33 mm and 26 mm for the anti-correlated painting.
In the beam commissioning, in order to measure the transverse beam distribution after painting,
a fast extraction scheme was designed by adjusting the extraction timing and extraction mode.
When the injection painting completed, the circulating beam was extracted quickly and then measured by a
multi-wire scanner which is on the beam transport line from the RCS to the target (RTBT). Therefore,
the transverse beam distribution just after painting can be obtained, as shown in Fig. 7.
It can be found that the transverse beam sizes increase with the painting ranges.
The fact that both of the horizontal and vertical beam sizes increase greatly with the horizontal
painting range suggests that the transverse coupling effect is serious.

\begin{figure}
\centering\includegraphics[width=1.0\textwidth]{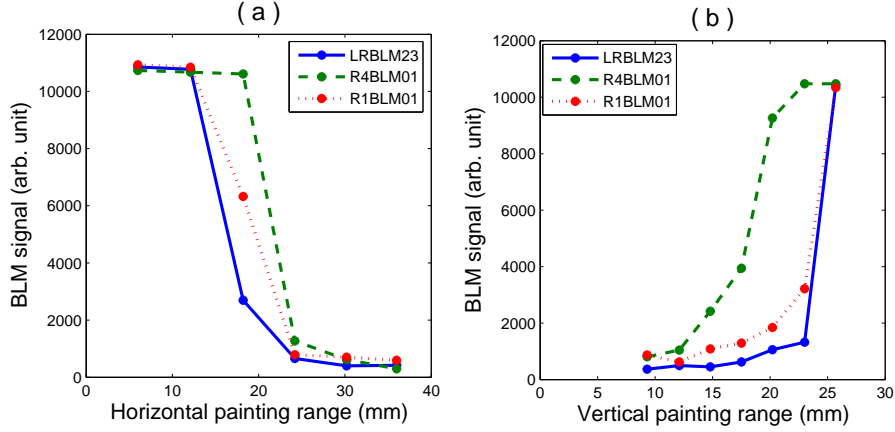}\\
\caption{Injection beam loss changing with the horizontal and vertical painting ranges
when the anti-correlated painting and working point (4.81, 4.87) were used.
Subgraph (a): the horizontal painting range was changed when the vertical
painting range was fixed to 12 mm; subgraph (b): the vertical painting range was changed when the horizontal
painting range was fixed to 27 mm.}
\end{figure}

By studying the variation of the injection beam loss with the transverse painting ranges, the painting ranges have also be optimized,
as shown in Fig. 8. In the figure, all the three beam-loss monitors (LRBLM23, R4BLM01, R1BLM01) are very close to
the injection point. LRBLM23 locates on the LRBT. R4BLM01 and R1BLM01 locate
on either side of the injection point of the RCS. These three beam-loss monitors
can basically reflect the injection beam loss. Therefore, it can be found that, when the horizontal
painting range is smaller than 27 mm, the injection beam loss begins to
grow rapidly. The reason of this phenomenon is that the circulating beam moves from the border to the center
in the horizontal painting process. If the horizontal painting range is too small, the circulating beam will be too close
to the border of the vacuum chamber and may cause injection beam loss.
When the vertical painting range is greater than 12 mm, the injection
beam loss begins to increase rapidly.
Combining the study of the transverse beam sizes and injection beam loss, the actual optimal horizontal painting range is 31 mm
which is close to the design horizontal painting range. The actual optimal vertical painting range is less than 15 mm which is very different
from the design vertical painting range.

\begin{figure}
\centering\includegraphics[width=1.0\textwidth]{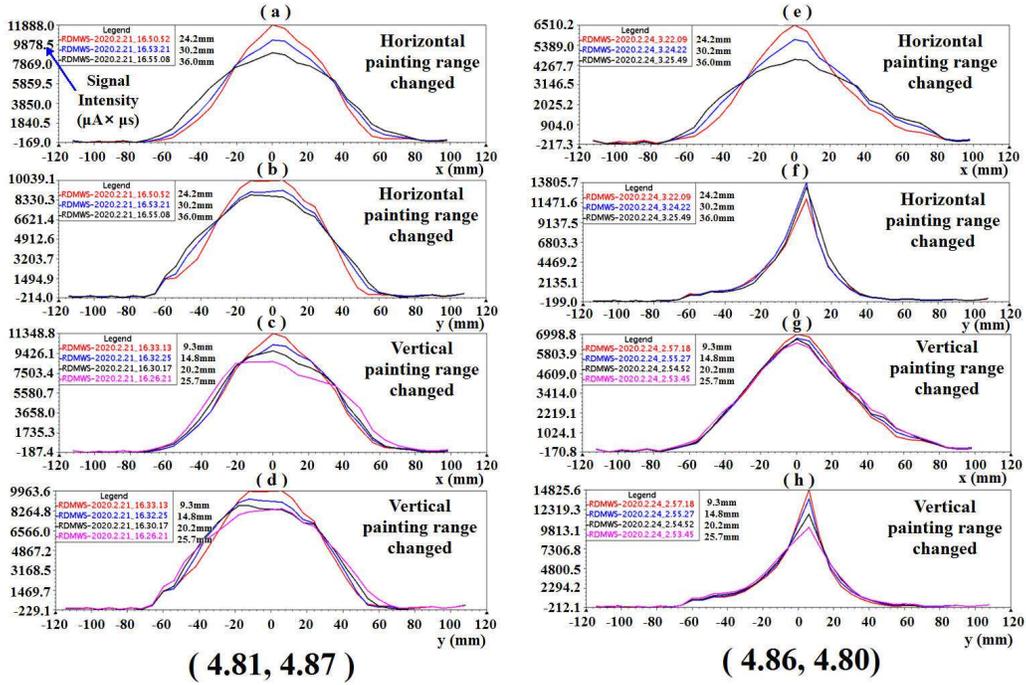}\\
\caption{Transverse beam distributions changing with the horizontal and vertical painting ranges for different working points
when the anti-correlated painting was used. Subgraphs (a), (b), (e) and (f): the horizontal painting range was changed when the vertical
painting range was fixed to 12 mm; subgraphs (c), (d), (g) and (h): the vertical painting range was changed when the horizontal
painting range was fixed to 27 mm.}
\end{figure}

According to the previous simulation results, the transverse coupling effect was very small for the anti-correlated
painting by using the nominal working point (4.86, 4.78). However, in the beam commissioning,
the transverse coupling effect was serious.
In order to study the transverse coupling effect, the anti-correlated painting processes for different working points
were studied. Fig. 9 shows the transverse beam distributions changing with the horizontal and vertical painting ranges for different working points.
It can be found that, with the working point (4.81, 4.87) which is used for the stable operation now, the transverse coupling effect is
serious. However, with the working point (4.86, 4.80) which is close to the design values, the transverse coupling effect is
small. Therefore, different transverse coupling effects caused by different working points result in the great difference
between the actual optimal and design vertical painting range.

Although the transverse coupling effect of the working point (4.86, 4.80) is small, the coherent oscillation effect caused by
this working point is very large which produces a large amount of beam loss.
On the contrary, for the working point (4.81, 4.87), although the transverse coupling effect is a little large,
the coherent oscillation effect is very small. By using the working point (4.81, 4.87),
the beam loss caused by the combination of these two effects is small
and can meet the requirement of the stable operation.
Therefore, in the beam commissioning, the transverse coupling effect and coherent oscillation effect should be considered
at the same time in choosing the optimal working point.

\subsection{Painting curves optimization}
\label{SecIII4}

\begin{figure}
\centering\includegraphics[width=0.9\textwidth]{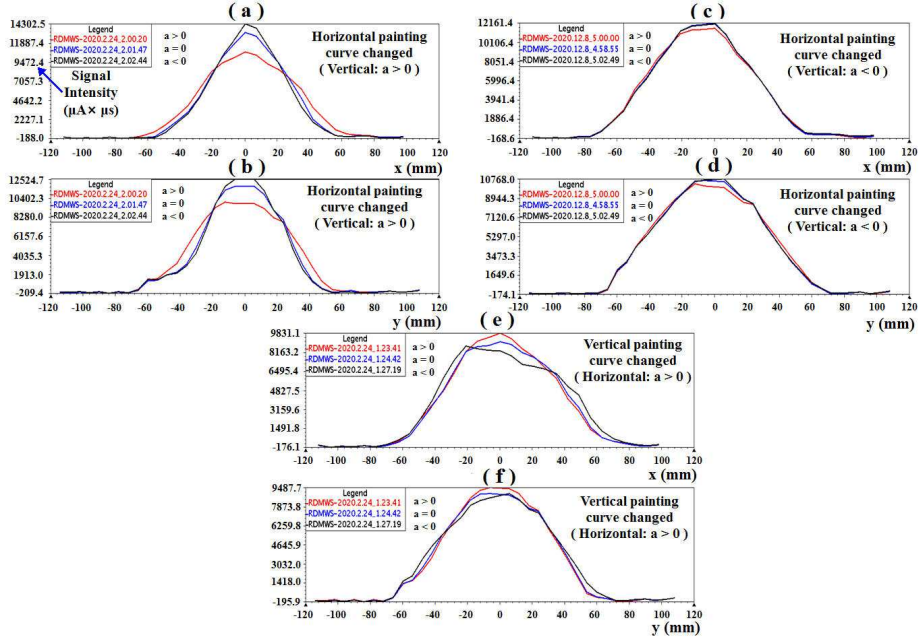}\\
\caption{Transverse beam distributions changing with the horizontal and vertical painting curves
when the anti-correlated painting and working point (4.81, 4.87) were used.
Subgraphs (a) and (b): the horizontal painting curve was changed when the vertical
painting curve was fixed to $a>0$; Subgraphs (c) and (d): the horizontal painting curve was changed when the vertical
painting curve was fixed to $a<0$; subgraphs (e) and (f): the vertical painting curve was changed when the horizontal
painting curve was fixed to $a>0$.}
\end{figure}

According to the simulation results of the design scheme,
when the painting curves with a high painting speed in the center and a slow painting speed in the border for both
horizontal and vertical directions, as shown in Fig. 4 and Eqs. (\ref{CX})-(\ref{CY}),
the transverse beam distribution is the most uniform and the injection results are the best after painting.

In the beam commissioning, by using the fast extraction scheme,
the transverse beam distribution just after painting
have be obtained. Fig. 10 shows the transverse beam distributions changing with the horizontal and vertical
painting curves.
$a=d^2I/dt^2$ is the variation rate of the slope of the painting curve.
It can be found that, when the vertical painting curve is fixed, the transverse beam sizes
after painting are minimal if the horizontal $a<0$. When the horizontal painting curve is fixed,
the transverse beam sizes after painting are minimal if the vertical $a>0$.
For the anti-correlated painting scheme of CSNS, the circulating beam is painted by the injection beam from the center to
the border in the horizontal plane and from the border to the center in the vertical
plane. Then, both the horizontal $a<0$ and vertical $a>0$ represent the painting curve with a slow
painting speed in the center and a high painting speed in the border.
Therefore, no matter in the horizontal or vertical plane,
when the painting curve with a slow
painting speed in the center and a high painting speed in the border, the transverse beam sizes
after painting are small.

\begin{figure}
\centering\includegraphics[width=0.65\textwidth]{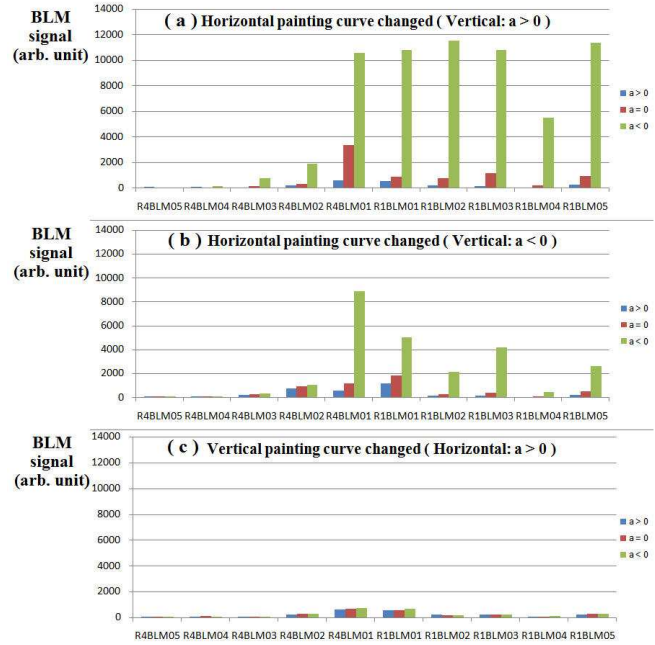}\\
\caption{Injection beam loss changing with the horizontal and vertical painting curves
when the anti-correlated painting and working point (4.81, 4.87) were used.
Subgraph (a): the horizontal painting curve was changed when the vertical
painting curve was fixed to $a>0$; Subgraph (b): the horizontal painting curve was changed when the vertical
painting curve was fixed to $a<0$; subgraph (c): the vertical painting curve was changed when the horizontal
painting curve was fixed to $a>0$.}
\end{figure}

\begin{figure}
\centering\includegraphics[width=0.5\textwidth]{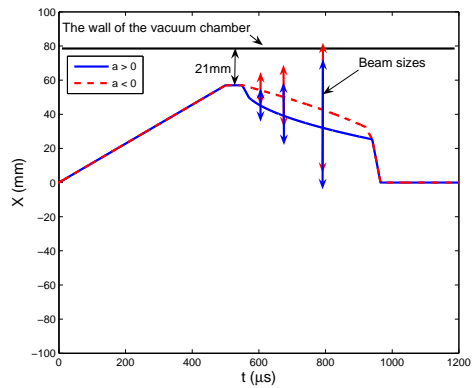}\\
\caption{The position relationship between the circulating beam and the wall of the vacuum chamber
at different times for different horizontal painting curves.}
\end{figure}

By studying the variation of the injection beam loss with the painting curves,
the painting curves have also be optimized. Fig. 11 shows the injection beam loss changing with the horizontal
and vertical painting curves. In the figure, all the ten beam-loss monitors (R4BLM05, R4BLM04, R4BLM03,
R4BLM02, R4BLM01, R1BLM01, R1BLM02, R1BLM03, R1BLM04,
R1BLM05) locate in the injection region.
It can be seen that, when the vertical painting curve is fixed, the injection beam loss
is minimal if the horizontal $a>0$. When the horizontal painting curve is fixed,
the injection beam loss is minimal if the vertical $a>0$.
In addition, when the horizontal painting curve with a slow
painting speed in the center and a high painting speed in the border (i.e. $a<0$), the injection beam loss is very large.
The main reason of this phenomenon is that the circulating beam is close to the wall of the
vacuum chamber of the injection system at the initial moment of the
horizontal painting. Fig. 12 shows the position relationship between the circulating beam and the wall
of the vacuum chamber at different times for different horizontal painting curves.
It can be found that, with the increase of the painting acceptance, the horizontal size of the circulating
beam also increases during the injection process which is caused by the space charge effects.
Compared to the high painting speed in the center (i.e. $a>0$), if the painting speed in the center is too slow (i.e. $a<0$),
the growing horizontal size of the circulating beam is more likely to exceed
the growing painting acceptance which may cause the injection beam loss in the horizontal plane,
as shown in Fig. 12. The simulation results also confirm the above explanation. Combining Figs. 10 and 11,
the actual optimal horizontal painting curve has a high
painting speed in the center and a slow painting speed in the border.
The actual optimal vertical painting curve has a slow painting speed in
the center and a high painting speed in the border.
Therefore, the actual optimal horizontal painting curve is consistent with the design horizontal painting curve
while the actual optimal vertical painting curve differs greatly from the design vertical painting curve.
Fig. 13 shows the difference between the design and actual optimal BV pulse current curves for the anti-painting scheme.
The main reason of the difference is caused by different working points used in the actual beam commissioning
and design scheme. The results of the machine study and simulation show that the actual optimal
painting curves (horizontal and vertical) are consistent with the design painting curves while
the same working point is used.

\begin{figure}
\centering\includegraphics[width=0.58\textwidth]{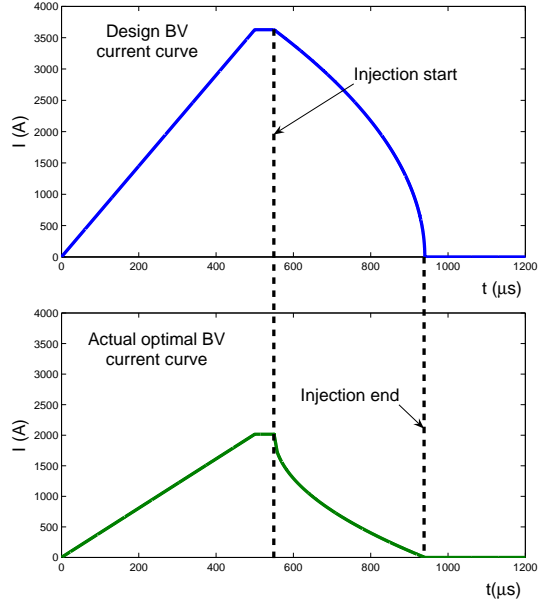}\\
\caption{The design and actual optimal BV pulse current curves for the anti-correlated painting scheme.}
\end{figure}

\subsection{Optimization of the extraction beam distribution}
\label{SecIII5}

\begin{figure}
\centering\includegraphics[width=0.45\textwidth]{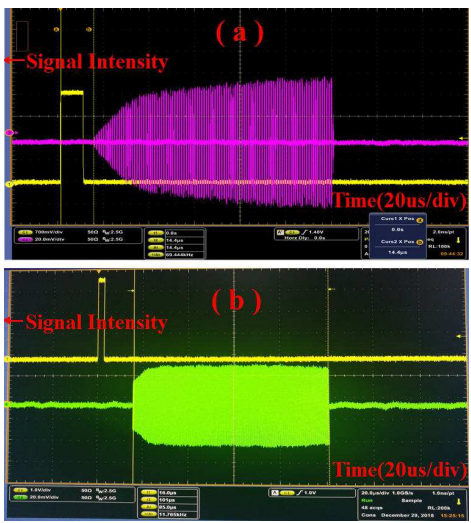}\\
\caption{The time structure of the injection beam. Subgraph (a): before cutting off the rising edge;
subgraph (b): after cutting off the rising edge.}
\end{figure}

Although the hollow painting method was not used,
the extraction beam distribution appeared a middle concave in the horizontal plane
which is very different from the simulation result of the design scheme.
There may be three reasons for this phenomenon: (1) the injection
timing was not accurate enough; (2) the matching between the injection beam and circulating
beam was not accurate enough; (3) the rising edge of the injection beam was of poor quality.

\begin{figure}
\centering\includegraphics[width=0.55\textwidth]{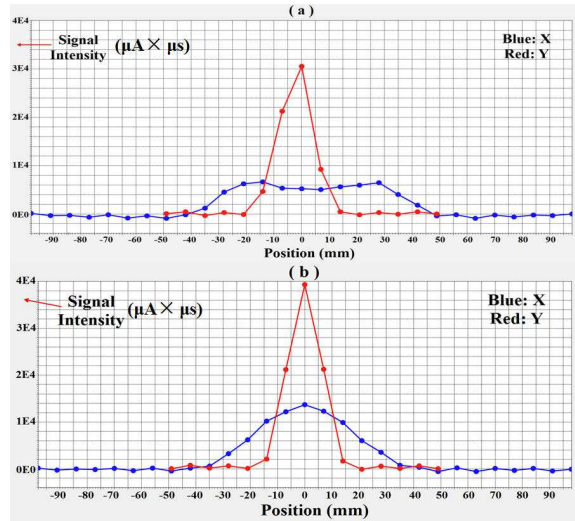}\\
\caption{Beam distribution on the target. Subgraph (a): before the injection optimization;
subgraph (b): after the injection optimization.}
\end{figure}

In the beam commissioning, the injection beam timing was precisely adjusted,
the injection beam and circulating beam were accurately matched,
and the rising edge of the injection beam was cut off by using the chopper.
Fig. 14 shows the time structure of the injection beam. The upper figure shows the
original injection beam and the lower figure shows
the injection beam after cutting off the rising edge.
After these three aspects of optimizations,
the extraction beam distribution in the horizontal plane
have been no longer concave in the middle, as shown in Fig. 15.
It is consistent with the simulation result which is close to the Gaussian distribution.

\subsection{Achieve 50 kW beam power on the target}
\label{SecIII6}

\begin{figure}
\centering\includegraphics[width=0.75\textwidth]{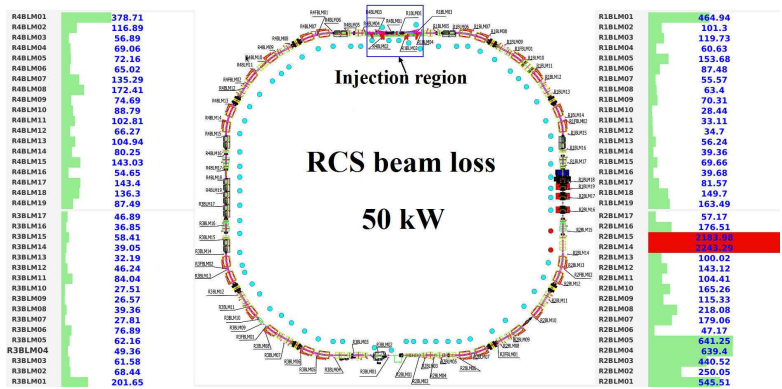}\\
\caption{RCS beam loss of 50 kW beam power on the target when the anti-correlated painting was used.}
\end{figure}

After the injection beam parameters matching, stripping foil optimization, phase space painting optimization,
extraction beam optimization and injection beam loss regulation, the injection beam loss has been well controlled
and the injection efficiency has been over 99\%. Combined with other aspects of the beam commissioning, the beam
power on the target has exceeded 50 kW and achieved the stable operation.
Fig. 16 and Fig. 17 show the RCS beam loss and beam distribution on the target of 50 kW beam power.
Fig. 18 shows the RCS DCCT display of 50 kW beam power on the target when the anti-correlated
painting was used. It can be seen from these figures that: both the injection efficiency and RCS transmission efficiency
are very high; the beam distribution and beam loss have met the requirements of the stable operation.

\begin{figure}
\centering\includegraphics[width=0.8\textwidth]{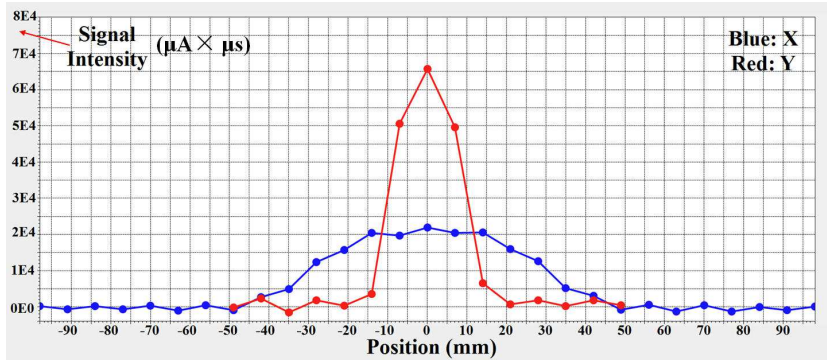}\\
\caption{Beam distribution on the target of 50 kW beam power when the anti-correlated painting was used.}
\end{figure}

\begin{figure}
\centering\includegraphics[width=0.75\textwidth]{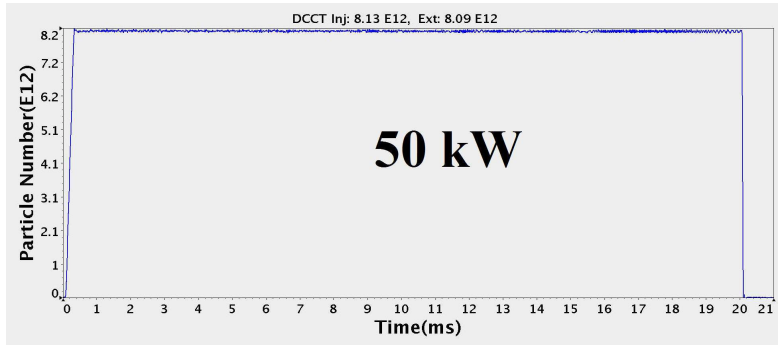}\\
\caption{RCS DCCT display of 50 kW beam power on the target when the anti-correlated painting was used.}
\end{figure}

\section{Summary and discussion}
\label{SecVI}

In this paper, firstly, the design scheme of the anti-correlated painting for CSNS/RCS was introduced and studied.
Then, the beam commissioning of the injection system was studied and discussed.
Compared to the simulation results of the design scheme,
the transverse beam distribution, transverse coupling effect and beam loss
of the actual anti-correlated painting were somewhat different.
By using the machine studies, we have analyzed these differences and studied their reasons.
After the optimization, the beam power on the target has exceeded
50 kW and achieved the stable operation.
In order to further improve the beam power on the target, the further comparative study between
the correlated and anti-correlated painting schemes should be conducted in the future.

\section*{Acknowledgments}
\label{Ack}

The authors would like to thank H.C. Liu, Y.L. Zhang, X. Qi,
P. Zhu, Z.H. Xu and other CSNS colleagues for helpful discussions.
M.Y. Huang also would like to thank Yoshiro Irie of KEK for helpful discussions.
This work was supported by National Natural
Science Foundation of China (Project Nos. U1832210 and 12075134).

\label{Ref}


\begin{thebibliography}{21}

\bibitem{Wang1} S. Wang et al., Introduction to the overall physics design of
CSNS accelerators, Chin. Phys. C, 33 (S2) (2009) 1-3.

\bibitem{Wei1} J. Wei, Synchrotrons and accumulators for high-intensity proton beams, Rev. Mod. Phys., 75 (4) (2003) 1383-1432.

\bibitem{Wei2} J. Wei et al., China Spallation Neutron Source - an overview of
application prospects, Chin. Phys. C, 33 (11) (2009) 1033-1042.


\bibitem{Huang1} M.Y. Huang, S. Wang, J. Qiu, N. Wang, S.Y. Xu, Effects of injection beam parameters
and foil scattering for CSNS/RCS, Chin. Phys. C, 37 (6) (2013) 067001.


\bibitem{Kawakubo1} T. Kawakubo, Analysis and measurement of beam dynamics in $H^-$ charge-exchange injection,
Nucl. Instrum. Methods Phys. Res. A, 265 (1988) 351-363.

\bibitem{Mohagheghi1} A.H. Mohagheghi et al., Interaction of relativistic $H^-$ ions with thin foils,
Phys. Rev. A, 43 (1991) 1345-1365.

\bibitem{Drozhdin1} A.I. Drozhdin, I.L. Rakhno, S.I. Striganov, L.G. Vorobiev,
Modeling multiturn stripping injection and foil heating for high intensity proton drivers,
Phys. Rev. ST Accel. Beams, 15 (2012) 011002.

\bibitem{Holmes1} J.A. Holmes, V.V. Danilov, J.D. Galambos, D. Jeon, D.K. Olsen,
Space charge dynamics in high intensity rings, Phys. Rev. ST Accel. Beams, 2 (1999) 114202.

\bibitem{Cousineau1} S. Cousineau, S.Y. Lee, J.A. Holmes, V. Danilov, A. Fedotov,
Space charge induced resonance excitation in high intensity rings,
Phys. Rev. ST Accel. Beams, 6 (2003) 034205.

\bibitem{Xu1} S.Y. Xu, S. Wang, Study on space charge effects of the CSNS/RCS, Chin. Phys. C, 35 (12) (2011) 1152-1158.

\bibitem{Wei3} J. Wei et al., Injection choice for Spallation Neutron Source ring, in: Proceedings
of the Particle Accelerator Conference, Chicago, United States, 2001, pp. 2560-2562.

\bibitem{Beebe-Wang1} J. Beebe-Wang, Y.Y. Lee, D. Raparia, J. Wei,
Transverse phase space painting for SNS accumulator ring injection, in: Proceedings
of the Particle Accelerator Conference, New York, United States, 1999, pp. 1743-1745.

\bibitem{Beebe-Wang2} J. Beebe-Wang, Y.Y. Lee, D. Raparia, J. Wei,
Beam properties in the SNS accumulator ring due to transverse phase space painting, in: Proceedings of the European Particle
Accelerator Conference, Vienna, Austria, 2000, pp. 1465-1467.

\bibitem{Tang1}  J.Y. Tang, J. Qiu, S. Wang, J. Wei, Physics design and study of the BSNS RCS injection system,
Chin. Phys. C, 30 (12) (2006) 1184-1189.

\bibitem{Gabambos1} J. Gabambos, J. Holmes, D. Olsen, ORBIT user's manual V1.0,
SNS-ORNL-AP Tech. Note 11, 1999.

\bibitem{Machida1} S. Machida, The Simpsons program user's reference Manual V1.0, July, 1992

\bibitem{Qiu1}  J. Qiu, J.Y. Tang, S. Wang, J. Wei, Studies of transverse phase space painting for
the CSNS RCS injection, Chin. Phys. C, 31 (10) (2007) 942-946.

\bibitem{weit1}  T. Wei, S. Wang, J. Qiu, J.Y. Tang, Q. Qin, Beam-loss driven injection optimization for CSNS/RCS,
Chin. Phys. C, 34 (2) (2010) 218-223.

\bibitem{Xu2} S.Y. Xu, S.X. Fang, S. Wang, The study of the space charge effects for RCS/CSNS, in: Proceedings
of HB2010, Morschach, Switzerland, 2010, pp. 420-424.

\bibitem{Huang2} M.Y. Huang et al., Measurement of the injection beam parameters by the multi-wire scanner for CSNS,
in: Proceedings of the 9th International Particle Accelerator Conference (IPAC2018), Vancouver, BC, Canada, 2018,
pp. 1014-1016.

\bibitem{Lu1} X.H. Lu, M.Y. Huang, S. Wang, Injection orbit matching for a rapid cycling synchrotron,
Phys.Rev. Accel. Beams, 21 (2018) 062802.

\end{thebibliography}
\end{document}